\documentclass[twocolumn]{aastex631}
\usepackage{natbib}
\usepackage{soul}
\usepackage{svg}
\usepackage{todonotes}
\usepackage{gensymb}
\usepackage{array,multirow}
\usepackage{ragged2e}
\usepackage{xcolor}
\usepackage{enumitem}
\setlist{itemsep=0.5em, topsep=0.2em} 

\newcommand{\mathtx}[1]{${\rm #1}$}
\newcommand{\Xh}[1]{${\rm X^{H}_{#1}}$}
\newcommand{\Th}[1]{${\rm T^{H}_{#1}}$}
\newcommand{\Xc}[1]{${\rm X^{#1}_{C}}$}
\newcommand{\Tc}[1]{${\rm T^{#1}_{C}}$}
\newcommand{\Yc}[1]{${\rm Y^{#1}_{C}}$}
\newcommand{\Xem}[1]{${\rm X^{#1}_{em}}$}
\newcommand{\Tem}[1]{${\rm T^{#1}_{em}}$}

\begin{document}

\title{Multi-modal encoder-decoder neural network for forecasting solar wind speed at L1}

\author[0000-0001-7927-2727]{Dattaraj B. Dhuri}
\affiliation{Center for Astrophysics and Space Science \\
New York University Abu Dhabi \\
PO Box 129188, Saadiyat Island, Abu Dhabi, UAE}

\author{Shravan M. Hanasoge}
\affiliation{Center for Astrophysics and Space Science \\
New York University Abu Dhabi \\
PO Box 129188, Saadiyat Island, Abu Dhabi, UAE}
\affiliation{Department of Astronomy and Astrophysics \\
Tata Institute of Fundamental Research \\
Mumbai, India}

\author{Harsh Joon}
\affiliation{Department of Astronomy and Astrophysics \\
Tata Institute of Fundamental Research \\
Mumbai, India}

\author{Gopika SM}
\affiliation{Department of Astronomy and Astrophysics \\
Tata Institute of Fundamental Research \\
Mumbai, India}

\author{Dipankar Das}
\affiliation{Parallel Computing Lab, Intel Labs, Bangalore, India}

\author{Bharat Kaul}
\affiliation{Parallel Computing Lab, Intel Labs, Bangalore, India}

\correspondingauthor{Dattaraj B. Dhuri}
\email{dbd7602@nyu.edu}



\begin{abstract}
The solar wind, accelerated within the solar corona, sculpts the heliosphere and continuously interacts with planetary atmospheres. On Earth, high-speed solar-wind streams may lead to severe disruption of satellite operations and power grids. Accurate and reliable forecasting of the ambient solar-wind speed is therefore highly desirable. This work presents an encoder-decoder neural-network framework for simultaneously forecasting the daily averaged solar-wind speed for the subsequent four days. The encoder-decoder framework is trained with the two different modes of solar observations. The history of solar-wind observations from prior solar-rotations and EUV coronal observations up to four days prior to the current time form the input to two different encoders. The decoder is designed to output the daily averaged solar-wind speed from four days prior to the current time to four days into the future. Our model outputs the solar-wind speed with Root-Mean-Square Errors (RMSEs) of 55~km/s, 58~km/s, 58~km/s, and 58~km/s and Pearson correlations of 0.78, 0.66, 0.64 and 0.63 for one to four days in advance respectively. While the model is trained and validated on observations between 2010 - 2018, we demonstrate its robustness via application on unseen test data between 2019 - 2023, yielding RMSEs of 53~km/s and Pearson correlations 0.55 for a four-day advance prediction. Our encoder-decoder model thus produces much improved RMSE values compared to the previous works and paves the way for developing comprehensive multimodal deep learning models for operational solar wind forecasting.
\end{abstract}



\section{Introduction} \label{sec:intro}
Magnetic fields dominate the dynamics of plasma in the solar atmosphere and corona. Coronal magnetic fields are constantly restructuring, thereby accelerating the surrounding plasma, resulting in a continuous emission of charged particles that form the ambient solar wind \citep{Bale2023}. The outward flow of solar wind carries the coronal magnetic field into the solar system and constitutes the heliosphere. The interaction of charged particles in the solar wind with the atmosphere and any extant magnetic field of the planets results in the formation of planetary magnetospheres. The variability of the solar wind is thus a primary driver of geomagnetic activity on Earth. Severe space weather is characterized by solar storms such as flares and coronal mass ejections (CMEs), which are typically accompanied by solar energetic particle (SEP) events. High solar wind plasma densities, speeds, and temperatures during these events are predominately responsible for major geomagnetic storms. High-speed streams of ambient solar wind are also known to cause significant heating of the atmosphere and sustained substorm activity, damage to satellite electronics and altering of their trajectories, and geomagnetic disturbances that can affect power grids \citep{Schrijver2014,Horne2018}. Accurate forecasting of the solar wind state is therefore important for maintaining the operations of modern technological infrastructure. 

The solar wind is accelerated in the outer solar corona and can change on timescales of minutes. It is primarily categorized as fast (400 - 800 km/sec) and slow (below 400 km/sec) \citep{McComas1995}. The fast wind is known to originate from coronal holes, whereas the slow wind possibly originates from the boundaries between these holes and low-altitude coronal streamer belts \citep{Hansteen2012}. Coronal holes (CH), with cooler plasma, are seen as dark areas in extreme ultraviolet (EUV) and soft X-ray images of the Sun and comprise open magnetic field lines that extend out to the heliosphere. CHs can be long lasting, sometimes for the duration of one full solar rotation $\sim$ 27~days, especially when solar magnetic activity is in a decreasing phase. In contrast, the slow solar wind is highly variable. Typically, the solar wind takes 2-4 days to reach L1, which NASA’s Advanced Composition Explorer (ACE) \citep{Stone1998} has been continuously monitoring since 1997. It takes only up to 30 min for the solar wind to arrive at Earth from L1. Therefore, knowledge of the solar wind at L1, 1-4 days in advance, may be greatly helpful in preparing for adverse consequences and potentially modeling the subsequent geomagnetic activity, e.g. \citep{Upendran2022}. High-resolution solar observations are now consistently available from state-of-the-art observatories like the Solar Dynamics Observatory (SDO) \citep{Pesnell-etall2012}. Solar wind forecasting efforts have focused on using such observations to achieve accurate forecasts up to 4 days in advance \citep{Yang2018,Upendran2020,Brown2022}.  

Although extreme variations in the solar wind are due to flares, CMEs, and SEPs, the background ambient solar wind changes are tied to the mechanism that generates it. The interaction of the fast solar wind with the surrounding slow solar wind forms coronal interaction regions (CIRs) that co-rotate with the Sun. These co-rotating CIRs give rise to the 27-day periodicity observed in the solar wind speed, density, and radial magnetic field \citep{Owens2013}. The CIRs play a particularly important role during the declining phase of the solar activity cycle due to which the 27-day persistence model for solar wind forecasting serves as a natural benchmark \citep{Owens2013}. The solar wind speed measured at a given time typically shows a high autocorrelation (${\rm \sim~0.75}$) with the wind measured up to one day prior. This correlation decreases and falls to 0.007 for the wind speed measured 4 days earlier \citep{Owens2013,Upendran2020}. Therefore, achieving accurate and reliable forecasting of the solar wind speed, particularly 2-4 days in advance, is challenging.

The physical processes responsible for the solar wind acceleration and variation are not completely known, and hence the success of physics-based numerical models in predicting the ambient solar wind has been limited. The National Oceanic and Atmospheric Administration (NOAA) provides solar wind predictions 1-4 days in advance using the semi-empirical WSA-ENLIL model \citep{Owens2008}. WSA-ENLIL (and other physics-based solar wind models such as MAS-ENLIL \citep{Schwenn2006} and Space Weather Modeling Framework \citep{Toth2005}) use surface magnetic-field observations compiled over one solar rotation (synoptic maps) and develop a simpler, potential-field source-surface (PFSS) model of the solar corona. The PFSS corona model is then used as an input to physics-based or empirical solar wind propagation models to simulate the ambient solar wind. These numerical models are computationally expensive and typically yield a correlation of 57\% with observations of solar wind speed, and show associated errors of $\sim$100 km/s \citep{Jian2015}, which are lower than the accuracy achieved by a persistence forecast one day in advance. Therefore, an improvement in existing solar wind forecasting is desirable.

The correlation between coronal holes and high-speed solar wind streams has been long known \citep{Krieger1973} and many studies have attempted to develop this correlation into an empirical forecasting model. These approaches have not been able to obtain a significant improvement over the purely physics-based approaches \citep{Upendran2020}. Alternatively,\citet{Yang2018} used attributes of the coronal magnetic field derived from a PFSS model in combination with the solar wind speed 27 days prior as input to a machine learning (ML) model based on neural networks. They reported a much improved performance with a correlation of 0.74 and an RMS error of 68 km/sec for forecasting an hourly averaged solar wind speed four days in advance. Their study motivated recent works dedicated to the development of ML-based models using deep learning neural networks \citep{Upendran2020, Brown2022}. Deep learning techniques have been highly successful in dealing with the image data for performing complicated tasks involving identification, regression, segmentation, object detection, etc. \citep{Goodfellow2016}. \citet{Upendran2020} used coronal EUV images using inception-based deep convolutional neural networks (CNN) \citep{GoogleNet2015} to obtain a correlation of 0.55 four days in advance. They also found the coronal hole features from the EUV images to be the most relevant for forecasting the fast wind. More recently \citet{Brown2022} improved on this study and used EUV images in combination with the 27-day prior solar wind speed as an input to attention-based neural networks and obtained an improved correlation of 0.63 with an RMSE of 72 km/sec for four-day advanced forecasting. Recently, \citet{Lin2024} used the PFSS magnetic-field maps as input to train a CNN to predict the hourly solar wind three to six days in advance with an average correlation coefficient of 0.53 and an average RMSE error of 81 km/sec.  The present work builds on the previous studies by \citet{Upendran2020} and \citet{Brown2022} and proposes an architecture that is naturally suited to exploit the long-term autocorrelation in the solar wind as well as the information in the latest available coronal EUV observations.

The 27-day periodicity of the solar wind autocorrelation is due to the CIR structures persisting over multiple solar rotations, particularly during quiet periods \citep{Owens2013}. Thus, known variations of the solar wind over the prior solar rotation, i.e., around the period of 27 days (and also 54 days, 81 days, etc.) prior, may be used as a basis to model variations at the present time. Motivated by neural natural language translation frameworks, we propose an encoder-decoder architecture that serves to ``time-translate" the solar-wind observations from the past solar rotations into a present-day prediction that includes past and current observations as well as forecasts for the next four days. The history of solar wind observations from past solar rotations is input to an encoder, while the ``time-translated" present- and future-day inferences are retrieved from a decoder. An additional encoder processes the recent time series of coronal EUV observations; thus, the proposed architecture incorporates two different modes of solar observations. We demonstrate that the proposed encoder-decoder model may be efficiently trained within a transfer learning set up, first using only history observations to achieve a first estimate of the future solar wind speeds and subsequently training the second encoder with recent coronal EUV observations while carrying forward the weights for the history encoder as well as the decoder. We show that the proposed architecture achieves a performance superior to the models of \citet{Upendran2020} and \citet{Brown2022}.

The present paper is organized as follows. In Section~\ref{sec:data_method}, we motivate the encoder-decoder architecture and present the details of the inputs to the model and the output. We also provide the details of the data used and the preparation of the training, validation, and test sets. We discuss the training stages and performance metrics used. In Section~\ref{sec:results}, we document the performance of our model and compare it with previous studies. We also compare the individual solar wind speed values predicted by the model with the true values. In Section~\ref{sec:discussion}, we summarize the work and performance of our model as well as comment on its significance. 

\section{Data and Method \label{sec:data_method}}
\subsection{Why encoder-decoder?}
\citet{Owens2013} prescribed a baseline solar-wind forecasting model using the characteristically high solar-wind autocorrelation with time delays of multiples of the approximate solar rotation period ${\rm \sim~27~days}$. This 27- (54, 81, and so on)-day autocorrelation arises from long-lived CIR structures, particularly during the quiet period of the solar cycle. While \citet{Upendran2020} did not incorporate the solar wind speed from the past solar rotations to exploit this autocorrelation, \citet{Brown2022} used only the value of the solar wind speed 27 days prior to the day for which the solar wind speed is to be predicted as an auxiliary input. The main inputs to both their models were the SDO/AIA EUV images for the current day (and up to four days prior for \citet{Upendran2020}). They demonstrated that deep learning networks using convolutional/spatial attention layers can extract features such as coronal holes and active regions that are relevant to the future solar wind speed. 
\begin{figure*}
    \centering
    \includegraphics[width=0.8\textwidth]{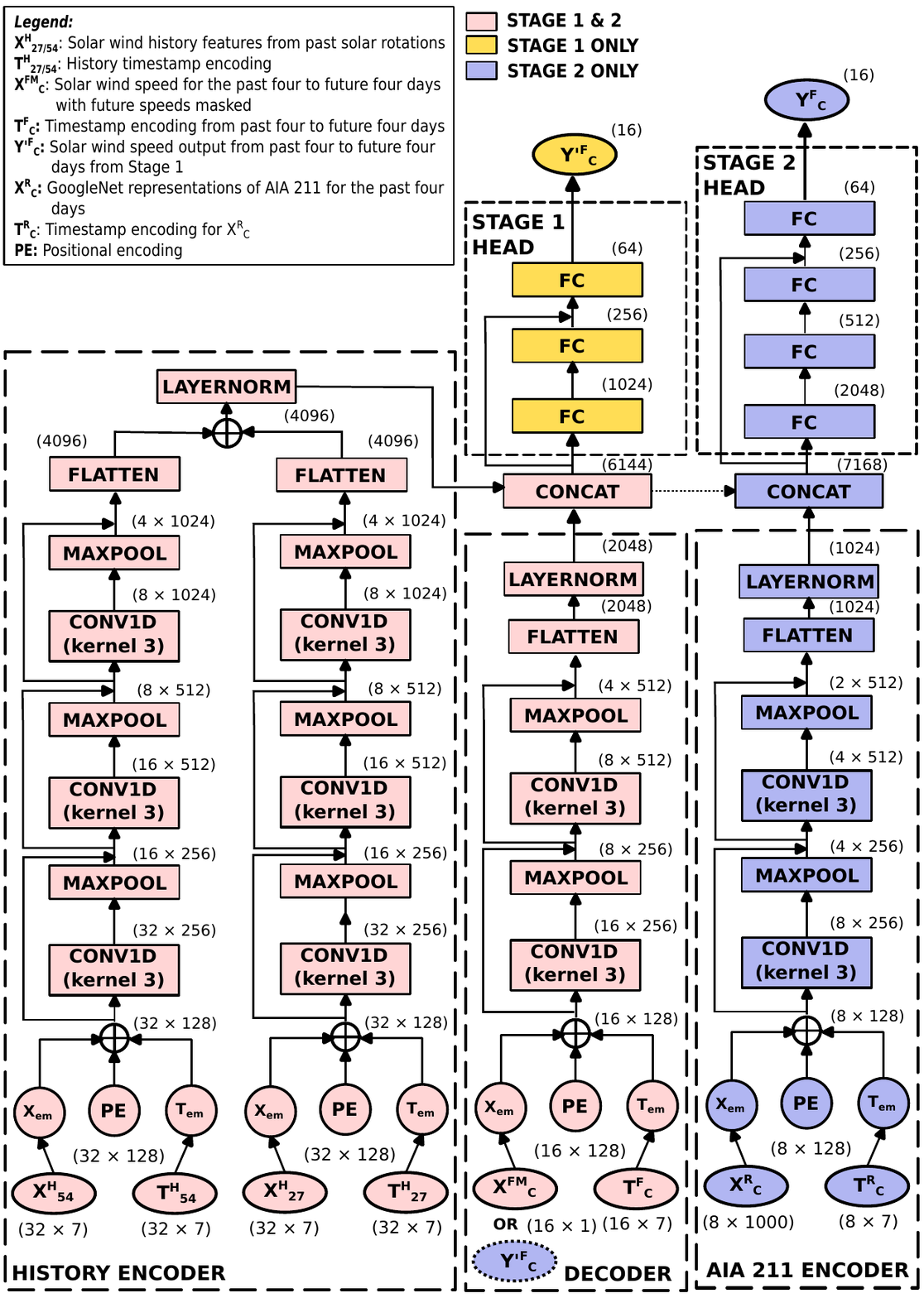}
    \caption{{\bf The multi-modal encoder-decoder architecture} used for predicting solar wind speed for four days into the future using solar wind history observed in the prior two solar rotations, the solar coronal EUV images from SDO/AIA \mathtx{211\AA} over the four days prior to current time, and the solar wind speed variation over four days prior to current time.}
    \label{fig:network}
\end{figure*}

Here, we use variations in the solar wind as observed during the past solar wind rotations as the basis on which to set up a sequence-to-sequence formulation for simultaneously forecasting the solar wind speed for the future four days. The formulation contains two input sequences: (a) the solar wind history from prior solar rotations as well as recent observations from four days prior (including the current day), and (b) recent observations of the solar corona from four days prior to the current day. The model outputs a sequence of the solar wind speed from four days prior to the current day to four days into the future. The sequence-to-sequence model is formulated using an encoder-decoder architecture \citep{Ilya2014,Chao2014} that has proven to be a cornerstone of highly sophisticated and successful models used for natural language processing, image captioning, generating text from audio, etc. The model architecture is shown in Figure~\ref{fig:network}. 

The model comprises two sets of encoders for processing two different modalities of the input data, and hence is a multi-modal architecture.
\begin{itemize}
    \item The first set --- the {\it History encoder} --- includes two identical encoders that process a time series of historical solar wind observations from the prior two solar rotations taken over a period of eight days centered around the 27th day and the 54th day before the current day.
    \item The second set --- the {\it AIA \mathtx{211\AA} Encoder} --- includes one encoder that processes a time series of the SDO/AIA EUV \mathtx{211\AA} taken from the observations of the past four days including the current day.
\end{itemize}  
The decoder is supplied with a record of solar wind speeds for the past four days, including the current day. These past values provide a predictive context for the decoder to predict future solar wind speed values via historical sequences. The decoder analyzes the abstract representations of historical solar wind trends from the past solar rotations, as well as changes in the EUV \mathtx{211\AA} over the previous four days. Then it outputs the solar wind speed for the past as well as four days into the future, yielding a forecast. The accuracy of the model in reconstructing the past solar wind speeds up to the current day helps to validate the architecture's effectiveness as large errors in the reconstruction would indicate defects in the encoder-decoder design e.g. insufficient feature extraction. The encoder-decoder formulation is thus set up to leverage the knowledge of solar wind variations (particularly CIRs) in prior solar rotations (in phase with the current time) and append it to predict changes based on the most recent solar activity, the state of the global coronal magnetic field as seen in the EUV \mathtx{211\AA} images, and the current trend in the solar wind speed.

\subsection{The history encoder}
This encoder set includes two identical encoder models, each processing the historical solar wind variation about 27 day and 54 days prior to the current time respectively. Wind measurements from two previous solar rotations are used to reinforce the expected long-term correlations from CIRs \citep{Owens2013}. Each history encoder takes a history of solar wind features (\Xh{}) from the ${\rm \pm~4~day}$ sequence centered around 27/54 days prior to the current time and hand-encoded timestamps (\Th{}) of these observations, as specified below.

\subsubsection{Timestamp encoding \Th{}}
Feature observations are \Xh{} taken over a period of 8 days centered at the 27th day and 54th day prior to the current day, denoted as \Xh{27} and \Xh{54}, respectively. The timestamps of these observations comprise day, month, year, and hour, since we are using daily averaged data on an hourly cadence. We also extend the notion of the timestamp to include the phase of the solar cycle during which these observations are taken, and the solar cycle number. These details of the observation time with respect to the solar cycle provide an additional context for the encoder to process information within \Xh{} since the long-term autocorrelation of the solar wind is known to depend on solar activity \citep{Owens2013}. Additionally, the timestamp includes the time in fractional days, relative to the observation time of the 27th/54th day prior to the current time. Each timestamp is normalized as follows.
\begin{enumerate}
    \item {\bf Hour}: Hour/24.0.
    \item {\bf Day}: Day/31.0.
    \item {\bf Month}: Month/12.0.
    \item {\bf Year}: (Year-1996)/50.0. Note that 1996 is the first year that we used the history data. The choice of 50.0 is arbitrary and may be retained for predictions until the year 2046.
    \item {\bf Relative day}: ${\rm (time - time_{27/54})}$/4.0. Varies between -0.9375 to 1.0 in steps of 0.0625 since we use the history observations every six hours, i.e. 32 observations within the eight days about the 27th/54th day.
    \item {\bf Solar cycle phase}: The time in years since the beginning of the solar cycle is normalized by the known duration of the solar cycle. For the current cycle \#25, we assume a duration of 11 years.
    \item {\bf Solar cycle number}: Assigned as 0.2, 0.4 and 0.6, respectively, for the cycles \#23, \#24 and \#25 considering an expected operational period of 50 years as mentioned above, i.e., approximately five solar cycles. This provides a context for variations between different solar cycles.
\end{enumerate}
Each \Th{27/54} includes this seven-entry timestamp encoding for the 32 timestamps specified above. 

\subsubsection{History features (\Xh{})}
\begin{enumerate}
    \item {\bf Solar Wind State} We use the hourly averaged solar wind data from OMNIWeb measured in situ at L1. The hourly averaged data since 1963 is publicly available at \url{https://omniweb.gsfc.nasa.gov/}. Here, we obtain daily averages sampled at an hourly frequency. Daily averages are used because, within a day, the solar wind speed usually does not show significant variability \citep{Upendran2020}; rather it is strongly autocorrelated \citep{Owens2013}. The hourly cadence also yields a sufficiently large number of samples necessary for training. Any short-duration variability within one day is primarily associated with SEPs and CMEs \citep{Owens2013}. The hourly moving averages for the daily period capture these variations to some extent.\\
    
    We require that a minimum number of 12 hourly observations be available for the moving daily average. We used the solar wind data from August 1996 onward (up to August 2024), as we found that the daily averages do not show any major temporal gaps due to missing data after this date since this corresponded to the commissioning of the NASA/ACE mission.\\

    The history data comprise the solar wind state defined by the daily averages of the speed, proton density, and magnetic field components \mathtx{B_x}, \mathtx{B_y}, \mathtx{B_z} obtained at an hourly cadence. All of these are obtained from the time \Th{27/54} specified above.

    \item {\bf Flare magnitudes} The hourly sampling of the daily average solar wind can still accommodate perturbations due to transient activity. These are the smoothed variations arising from SEPs and CMEs, which in turn are associated with flares. By including records of flare activity, we assign a context to the observed solar-wind history during the past two CRs. Flare records are issued by GOES in terms of flare classes and may be accessed via SunPy \citep{sunpy}. Flare classes A, B, C, M and X represent magnitudes of the X-ray flux recorded near the Earth on a logarithmic scale. We assign a flare magnitude to each of the flare categories starting from \mathtx{10^0} for the A class and increasing to \mathtx{10^5} for the X class in powers of 10. The flare magnitudes within an hour are summed up and averaged (by the duration of the hour) to represent a value for the flare activity context. All of these are also obtained from the time \Th{27/54} specified above.

    \item {\bf Sunspot number} \citet{Owens2013} noted that long-duration persistence is closely tied to solar activity, and we therefore include the daily sunspot number as an additional context for solar wind history over the past two CRs. The daily sunspot number data are readily available from SILSO (\url{https://www.sidc.be/SILSO/home}).
\end{enumerate}
The \Xh{27/54} history time series of solar wind features thus includes a total of seven features for the 32 timestamps specified above. 

\subsubsection{Architecture and operation}
The two history encoders are identical and are as shown in Figure~\ref{fig:network}, processing as inputs the separate history features \Xh{27/54} and the corresponding timestamp encoding \Th{27/54}. The encoder architecture comprises embedding layers, a series of 1D convolution blocks followed by a flattening layer, giving a representation of the encoded history data. The details of these layers and their operations are as follows. 
\begin{itemize}
    \item {\bf Embedding layers and positional encoding}: Embedding layers are used in both the history and the AIA encoder to transform the input into representations of a dense vector space. These are 1D convolution layers with kernel size 1 and stride 1 that serve to project the input as an abstract representation. There are two embedding layers \Xem{} and \Tem{}, each for \Xh{27/54} and \Th{27/54}, respectively. Separate positional encoding input PE, common to sequence-to-sequence models \citep{Vaswani2017}, is also included. Outputs of \Xem{}, \Tem{}, and PE undergo a layernorm operation and are subsequently added together and passed on to the next layer. The outputs of the embedding layers and the PE are 128-dimensional. The increase in dimensionality from 7 to 128 allows learning of any existing mutual relationships within the History features \Xh{} as well as the timestamp encodings \Th{} into the corresponding learned abstract representations of the embedding layer. 
    \item {\bf 1D Convolution block}: A series of 1D convolution blocks follow the embedding layers. Each convolution block comprises a 1D convolution layer, a Maxpooling layer, and a residual connection. The 1D convolutional layer is made up of convolutional filters with a kernel size of 3 and stride 1 that perform convolutions along the time dimension of the inputs. This is followed by a Maxpooling layer. A residual connection projects the inputs to match the shape of the output from the Maxpooling layer and adds it to the Maxpooling output. We use three such convolution blocks that convert the initial ${\rm 32~\times~128}$ output from the embedding layer to an encoded representation of ${\rm 4~\times~1024}$.
    \item {\bf Flatten, addition and Layernorm}: The output from the convolution blocks is flattened into a one-dimensional vector. The flattened output from both the encoders are added and fed into a Layernorm layer and passed on to the next stage.
    \item {\bf Activation function}: Each 1-D convolution layer uses a {\it relu} activation function.
    \item {\bf Weight initialization}: The weights and biases of the convolution layers are initialized using the default weight initialization scheme in PyTorch. 
\end{itemize}
    
\subsection{The AIA Encoder}
The AIA encoder encodes the information from the SDO/AIA EUV images of the solar corona. We used EUV \mathtx{211\AA} images as it is shown to yield the best results in the works of \citet{Upendran2020} and \citet{Brown2022}.

The SDO/AIA instrument provides observations of the solar atmosphere and corona in UV and EUV wavelengths. These observations are available since May 2010 at a high resolution of \mathtx{4k \times 4k} and a very high cadence of 2 secs. \citet{Galvez_2019} prepared an ML-ready dataset from SDO observations, called SDOML, that includes a standardized AIA dataset sampled at a cadence of 6 min and on a spatially reduced \mathtx{512 \times 512} grid. We use the SDOML AIA images from the EUV channel \mathtx{211\AA} for this study. 

\subsubsection{Current EUV \mathtx{211\AA} image GoogleNet representations \Xc{R}}
Deep-learning encoder-decoder architectures typically require a large number of samples (\mathtx{\sim 1M}) in order to be adequately trained. However, in our case, even after using the daily averaged data sampled at an hourly cadence, the size of the training dataset is only \mathtx{\sim 50k}. Therefore, following \citet{Upendran2020,Brown2022}, we use a pre-trained GoogleNet \citep{GoogleNet2015} to first extract morphological features relevant for developing an empirical relationship for the solar wind. \citet{Upendran2020} showed that the GoogleNet features extracted in this manner are meaningful and can capture the appropriate CH and AR regions from the data. 

For obtaining GoogleNet representations, we follow the preparation of the EUV images according to \citep{Brown2022}. Consequently, we first crop the central \mathtx{300 \times 300} portion of the original image \mathtx{512 \times 512} and rescale it to \mathtx{224 \times 224} to match the required standard image size. The dynamic range of the AIA pixels in SDOML is large and following \citet{Upendran2020} and \citet{Brown2022}, we clip the pixel values in the range 25 and 2500 and subsequently use their natural logarithms. These are then converted to the RGB format required by GoogleNet and scaled according to the standardization requirements of the GoogleNet dataset. 

The AIA encoder is configured to process the \mathtx{211\AA} GoogleNet representation thus obtained within the prior four days from the current time, prepared as a time series \Xc{R} with corresponding timestamps \Tc{R}. The preparation of the timestamp is explained in detail in the next section.

\subsubsection{Timestamp encoding \Tc{R}}
Unlike the solar-wind data, the SDO/AIA may occasionally have missing observations. However, we obtain \Xc{R} considering all possible observations that are available within the past four days of the current time and selecting the eight observations from those that are adequately spaced apart in time. Our goal is to ideally sample the EUV observations at a cadence of 12 hours within the past four days, thus eight observations. Allowing for the missing data, we sample any eight observations within this period which are within 12 hours of the ideal time, as well as are separated by at least six hours between the consecutive selected observations. After screening all available observations that satisfy these criteria, we hand-encode the corresponding timestamps identical to the history encoder. Note that the relative day in this case varies between -0.875 to 0, with eight values, although not necessarily always in steps of 0.125.

\subsubsection{Architecture}
The AIA encoder architecture is identical to the history encoder, except that only two convolution blocks are used. Here, the 128-dimensional embedding layer is used to first compress the 1000-dimensional GoogleNet representations \Xc{R} of the current EUV \mathtx{211\AA} images. The output from the final convolution block is of shape ${\rm 2~\times~512}$ that is flattened and passed through a layernorm layer. This represents an encoded representation of the EUV \mathtx{211\AA} time series from the prior four days and is passed on to the next stage.

\subsection{Decoder}
\subsubsection{Timestamp encoding \Tc{F} and Future Value Masked Current Solar Wind Speed \Xc{FM}}
The decoder is fed a time series of solar wind speeds taken from within the past four days from the current time \Xc{FM} and the corresponding encoded timestamps \Tc{F}. \Tc{F} also includes the future four days' timestamps for which speeds are to be predicted. The corresponding future values of the solar wind speeds in the input are masked by assigning values of 0.0. These are sampled at intervals of every 12 hours, i.e., a total of 16 timestamps, including the history and future. Thus, future values are obtained at a frequency of every 12 hours for the following four days. The timestamp encoding for these is identical to that of the history encoder. In this case, the values of the relative day vary from -0.875 to 1.0 in increments of 0.125.

\subsubsection{Architecture}
The architecture has a structure identical to the history encoder with only two 1D convolution blocks as shown in Figure~\ref{fig:network}. The output of the convolution layer is an array of shape ${\rm 4~\times~512}$ that is flattened and passed through a Layernorm operation to yield a representation of the current solar wind speed. This will be used to decode encoded information from the history and the AIA encoders in the subsequent stage.

\subsection{Training}
The encoder-decoder model uses two modalities of the data -- the solar wind history and the SDOML/AIA \mathtx{211\AA} image GoogleNet representations. We use the solar wind history data from August 1996 onward i.e. for three solar cycles \#23, \#24 and \#25. The SDOML/AIA data are available only from May 2010 onwards, i.e., for solar cycles \#24 and \#25 only. Hence, we train the model in two stages. In the first stage, only the history encoder is trained along with the decoder using the data from 1996 onward. In the second stage, the AIA encoder is also included, and training is performed using observations from May 2010 onward. The weights of the history encoder and the decoder, learned during the first stage, are retained and carried over to the next stage. Additionally, the predicted output values of the solar wind speed in the first stage \Yc{'F} are used as input to the decoder during the second stage instead of future masked values. The details of the data splitting for training, validation and test during both the stages, the architecture and the operation of the decoder heads for obtaining the output, the training process, and performance metrics are explained below in detail.

\begin{figure*}
    \centering
    \includegraphics[width=0.8\textwidth]{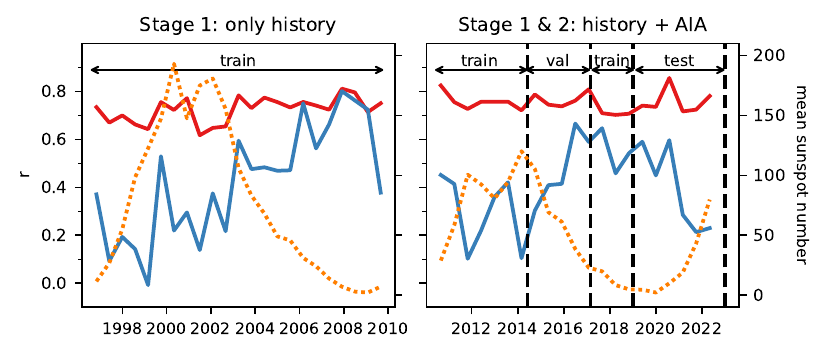}
    \caption{{\bf Variation of the solar wind speed auto-correlation.} The values of the Pearson correlation (left Y-axis) for the one- and the 27-day persistence models are plotted versus solar cycle \#23 and \#24 (the sunspot number is shown on the right Y-axis). Note that both one- and 27-day persistence scores drop during the solar-activity maximum and attain peak values during solar minima. The maximum for each solar cycle is marked by the dashed lines. The persistence scores are obtained using the solar wind speed values every six months. The training and validation data are selected from cycles \#23 and \#24 considering these biases as shown.}
    \label{fig:data_parting}
\end{figure*}

\subsubsection{Data splitting for training, validation and test}
Solar wind speed and its evolution are highly variable in the solar cycle. The solar-wind autocorrelation and the long-term (e.g., 27-day) persistence correlation is highest during the quiet phases of the solar cycle, as shown in Figure~\ref{fig:data_parting}, while during the active phases the long-term correlation is significantly reduced.  Using data from only quiet or active phases of the solar cycle for validation or test sets biases the performance scores and training.  In particular, the validation/test data must not include samples from only the quiet phase that naturally have a high baseline persistence correlation. \citet{Upendran2020} and \citet{Brown2022} have used a five-fold cross-validation approach to train the model using five different training/validation splits from the solar cycle \#24. Here, we instead actively consider biases that arise in data splitting due to solar activity phases and split the data to include samples from the quiet and the active phases of cycle \#24, in both the training and validation sets. The robustness of performance is ensured using completely unseen test data sets that comprise observations from cycle \#25. Cycle \#23 predates the SDO/AIA era and, therefore, is only included for training the history encoder during Stage 1 described in the following. Details of the data splitting are summarized in Table~\ref{tab:data_splitting}.

\begin{table*}
\caption{{\bf Splitting the data into training, validation and test.} Note that the number of samples in each set is reduced from Stage 1 to Stage 2 because of missing data in SDO/AIA.}
\centering
\begin{tabular}{l c c}
\hline
   & {\bf Stage 1} & {\bf Stage 2} \\
   & (History Encoder Only) & (History \& AIA Encoder) \\
\hline
  & 01 Aug 1996 -- 12 May 2010 &  --- \\
 {\bf Training} &  \multicolumn{2}{c}{\& 13 May 2010 -- 31 Dec 2013}    \\
 & \multicolumn{2}{c}{\& 05 Mar 2017 -- 31 Dec 2018} \\
 & Total samples \# : 131272 & Total samples \# :  33176 \\
\hline
{\bf Validation}  & \multicolumn{2}{c}{05 Jan 2014 -- 28 Feb 2017}      \\
 & Total samples \# : 22788 &  Total samples \# : 19735 \\
\hline
{\bf Test}  &  \multicolumn{2}{c}{05 Jan 2019 -- 01 Aug 2024}    \\
 & Total samples \#: 30528 & Total samples \#: 24182 \\
\end{tabular}
\label{tab:data_splitting}
\end{table*}
\begin{figure*}
    \centering
    \includegraphics[width=0.8\textwidth]{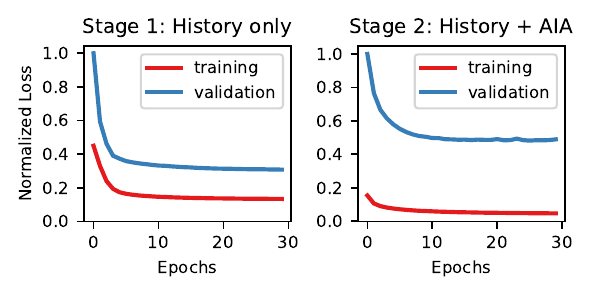}
    \caption{{\bf Progression of losses during training.} The loss function systematically decreases in both the training and validation phases in successive epochs, implying a smooth convergence during training.}
    \label{fig:Losses}
\end{figure*}

\subsubsection{Training stages}
The training progresses in two stages. In the first stage, the decoder is connected only to the history encoder. In the second stage, both the history and the AIA encoders are connected to the decoder. In both stages, the output of the encoder and decoder is concatenated and passed through a fully connected network head (FC head). The details of the input and output of the head, its architecture, and operation are as follows.

\paragraph{Stage 1 FC head:}
\begin{itemize}
    \item {\bf Inputs and outputs:} The inputs to the stage-1 FC head are vector representations from the history encoders and the decoder. The stage-1 FC head outputs the solar wind speeds from four days in the past to four days into the future \Yc{'F}. The speeds are predicted at a cadence of every 12 hour and hence the output has a size of 16.
    \item {\bf Architecture and operation:} The history encoder outputs a vector representation of size 4096 and the decoder outputs a vector representation of size 2048. These are concatenated with a Concat layer and passed to the Stage 1 head comprising three FC layers that ultimately reduce the representations to the desired output size of 16. A residual skip connection is also included after the Concat layer that feeds into the output from the penultimate FC layer. The FC and output layers all use sigmoid activation. The weights are initialized using the default scheme in PyTorch.
\end{itemize}

\paragraph{Stage 2 FC head:}
\begin{itemize}
    \item {\bf Inputs and outputs:} Inputs to the Stage-2 FC head are vector representations from the history encoders, the AIA encoder and the decoder. The predicted speeds for the future four days from Stage 1 form the input to the decoder for training during Stage 2. The Stage-2 FC head also yields the final predicted solar-wind speeds from four days in the past to four days in the future \Yc{F}. The speeds are predicted at a cadence of every 12 hours, and hence the output has a size of 16.  
    \item {\bf Architecture and operation:} The history encoders output vectors of size 4096, the AIA encoder outputs a vector of size 1024, and the decoder outputs a vector representation of size 2048. These are concatenated with a Concat layer and passed to the Stage 2 head comprising four FC layers that ultimately reduce the representations to the desired output size of 16. Similarly to the stage 1 head, a residual skip connection is also included after the Concat layer that feeds into the output from the penultimate FC layer. The FC and the output layers all use Sigmoid activation. The weights for the AIA encoder and the stage 2 head are initialized using the default scheme in PyTorch. For the history encoders and decoder, the weights are carried forward from the Stage 1 training. The history encoders' weights for the first eight epochs are frozen, and only the AIA encoder weights are updated. Subsequently, all weights are updated during training.
\end{itemize}

\begin{table*}
\caption{{\bf Solar wind prediction performance.} The model predicts the outputs from the prior four days to four days into the future every 12 hours. The Pearson correlations (r) and RMSEs for these predictions are listed for both Stage 1 and Stage 2 outputs. Note that the model predictions for the days -3 to 0 are only a passive reconstruction of the historical sequence input to the decoder and does not involve predictive logic. The true model predictions are obtained for days 1 to 4.}
\centering
\begin{tabular}{c|c c c c|c c c c}
\hline
 Day  & \multicolumn{4}{c|}{\bf Stage 1} & \multicolumn{4}{|c}{\bf Stage 2} \\
 Relative to  & \multicolumn{4}{c|}{(History Encoder Only)} & \multicolumn{4}{|c}{(History \& AIA Encoder)} \\
 Current time  & \multicolumn{2}{c}{\bf Validation} & \multicolumn{2}{c|}{\bf Test} & \multicolumn{2}{|c}{\bf Validation} & \multicolumn{2}{c}{\bf Test} \\
 \hline
 & r & RMSE & r & RMSE & r & RMSE & r & RMSE \\
 &  & (km/s) &  & (km/s) &  & (km/s) & r & (km/s) \\
\hline
 & \multicolumn{8}{c}{Past and Current} \\
 \hline
 -3 & 0.99 & 12.99 & 0.99 & 12.33 & 0.97 & 26.55 & 0.96 & 26.23 \\
 -2 & 0.97 & 16.87 & 0.97 & 15.67 & 0.95 & 27.02 & 0.94 & 26.64 \\
 -1 & 0.97 & 15.90 & 0.97 & 14.79 & 0.94 & 28.58 & 0.94 & 30.72 \\
  0 & 0.98 & 15.26 & 0.98 & 13.31 & 0.94 & 37.90 & 0.95 & 33.78 \\
\hline
 & \multicolumn{8}{c}{Future} \\
 \hline
  1 & 0.82 & 44.50 & 0.80 & 39.43 & 0.78 & 54.67 & 0.78 & 46.88 \\
  2 & 0.62 & 60.30 & 0.58 & 53.21 & 0.66 & 58.30 & 0.61 & 50.89 \\
  3 & 0.55 & 63.25 & 0.51 & 56.31 & 0.64 & 58.17 & 0.57 & 52.48 \\
  4 & 0.56 & 63.25 & 0.48 & 57.99 & 0.63 & 57.60 & 0.55 & 53.25 \\
\hline
\end{tabular}
\label{tab:perform1}
\end{table*}

\begin{table*}
\caption{{\bf Comparison of solar wind prediction performance}. The performance of encoder-decoder model of this work has been compared with earlier studies \\citet{Upendran2020,Brown2022}. \citet{Upendran2020} used four different models to obtain predictions for the next four days while \citet{Brown2022} obtained predictions only for the fourth day into the future. Both these models used validation data between 2010 and 2018, albeit with five different cross-validation folds. In contrast, this works uses validation data that is balanced for the dependence of the baseline persistence score and the solar-wind autocorrelation on solar activity.}
\centering
\begin{tabular}{c|c c | c c | c c | c c}
\hline
  & \multicolumn{2}{c|}{This Work: Stage 1 } & \multicolumn{2}{c|}{This Work: Stage 2} & \multicolumn{2}{c|}{\citet{Upendran2020}} & \multicolumn{2}{c}{\citep{Brown2022}} \\
 \hline
 Future & r & RMSE & r & RMSE & r & RMSE & r & RMSE \\
 Day &  & (km/s) &  & (km/s) &  & (km/s) & r & (km/s) \\
\hline
  4 & 0.56 & 63.25 & 0.63 & 57.60 & 0.54 & 81.21 & 0.63 & 72.21 \\
  3 & 0.55 & 63.25 & 0.64 & 58.17 & 0.55 & 80.28 & --- & --- \\
  2 & 0.62 & 60.30 & 0.66 & 58.30 & 0.52 & 83.06 & --- & --- \\
  1 & 0.82 & 44.50 & 0.78 & 54.67 & 0.54 & 80.27 & --- & --- \\
\hline
\end{tabular}
\label{tab:perform2}
\end{table*}

\subsubsection{Loss function and hyperparameter tuning}
We use the mean squared error loss function. The distribution of solar wind speeds is not uniform, and extreme values at both ends occur very infrequently in comparison to values in the middle ranges (see Appendix 2). We experimented with two approaches to counter this sample imbalance --- oversampling and weighting the mean squared error loss. We found that the latter approach, i.e. weighting the mean squared loss, provided optimum results. We found that scaling the losses by a constant factor (a hyperparameter) of 100 times the normalized solar wind speed value in the output yields the best result. Other hyperparameters for training, such as learning rate, batch size, and dropout (used to mitigate overfitting), were also fine-tuned to obtain optimal predictions on the validation data. The weights are L2 regularized to further limit overfitting. Although the model takes a few hours to train on a GPU, once trained, the inference is achieved extremely fast $\sim$ 10 ms on a modern computer e.g. a single core of Intel(R) Xeon(R) CPU E5-2620 v4 @ 2.10GHz.

The resultant training progression of the loss function values is shown in Figure~\ref{fig:Losses}. Losses for both training and validation decrease in successive epochs, implying smooth convergence during training. However, the final loss function value for validation is higher than training, indicating a residual bias in the model due to some overfitting.  

\subsubsection{Performance metrics}
Following earlier studies related to the prediction of solar wind using deep learning, and in particular \citet{Upendran2020} and \citet{Brown2022}, we use the Pearson correlation (r) between the true and predicted solar wind speed, as well as the root mean squared error (RMSE) to evaluate the performance of the model. These are evaluated on all datasets --- training, validation and test --- and compared with the previous studies. Additionally, we visualize the true and predicted solar wind speeds to identify the biases in the model.

\section{Results \label{sec:results}}
\begin{figure*}
    \centering
    \includegraphics[width=\textwidth]{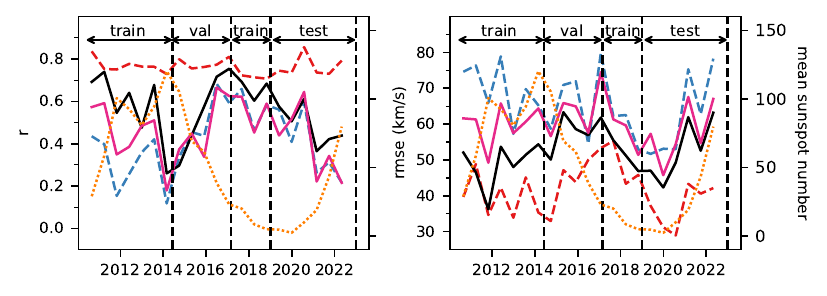}
    \caption{{\bf Variation of the Pearson correlation (r) and RMSE for the predicted solar wind speed for fourth day in future as a function of the sunspot number.} The left and the right panel shows the r and RMSE values respectively across the Solar Cycle 24 and 25 considered for the evaluation of the solar wind speed prediction. The sunspot number is indicated by the yellow dotted curve. The one-day and 27-day persistence baseline values are indicated by the dashed red and blue lines respectively. The Stage 1 and Stage 2 model values are indicated by the solid magenta and black lines respectively. Our models, particularity after Stage 2, consistently yield higher r values and significantly lower RMSE values compared to the 27-day persistence baseline.}
    \label{fig:resultsCorrVar}
\end{figure*}
\subsection{Performance Scores}
The correlation r and RMSE values for Stage 1 and Stage 2 for the past and future four days are listed in Table~\ref{tab:perform1}. For the past four days, we expect the correlations to be very high and the RMSEs to be relatively low because the solar wind speed values for these days are also input to the decoder. Nevertheless, this accurate reconstruction of the past solar wind speed fed into the decoder thus validates the encoder-decoder architecture design. For Stage 1, the r values are $> 0.97$ and the RMSEs are ${\rm \sim~15~km/s}$ for all the past days, both for validation and for the test. The r values for the one-day future forecast are also high, 0.82 and 0.80 for the validation and test respectively, and the corresponding RMSEs are ${\rm \sim 40~km/s}$. As forward-looking time increases, the r values decrease and the RMSE values increase, which is as expected. For the four-day future forecast, the r values are 0.56 and 0.48 for the validation and test data respectively, and the RMSE values are 63.25 km/s and 58 km/s, respectively. The Stage 1 encoder uses only the history data and both the r and RMSE values decrease significantly after one day into future, similar to persistence. 

For Stage 2 that uses the AIA EUV \mathtx{211\AA} data in addition to the solar wind history, the r and RMSE scores are less than those in the Stage 1 for past predictions as well as for one-day into the future. The performance scores, however, improve significantly for predictions two days into the future onward. The improvement beyond the second day compared to the history (as well as the persistence autocorrelation) is due to the model's ability to detect patterns in AIA images such as coronal holes and active regions \citep{Upendran2020} that can be related to the mechanisms that generate the solar wind. The Stage 2 model yields r values of 0.63 and 0.55 and RMSE values of 57.60 km/s and 53.25 km/s on the validation and test data, respectively, for a prediction four days into the future.

\begin{figure*}
    \centering
    \includegraphics[width=0.75\textwidth,trim={1cm 2.1cm 0.2cm 2cm},clip]{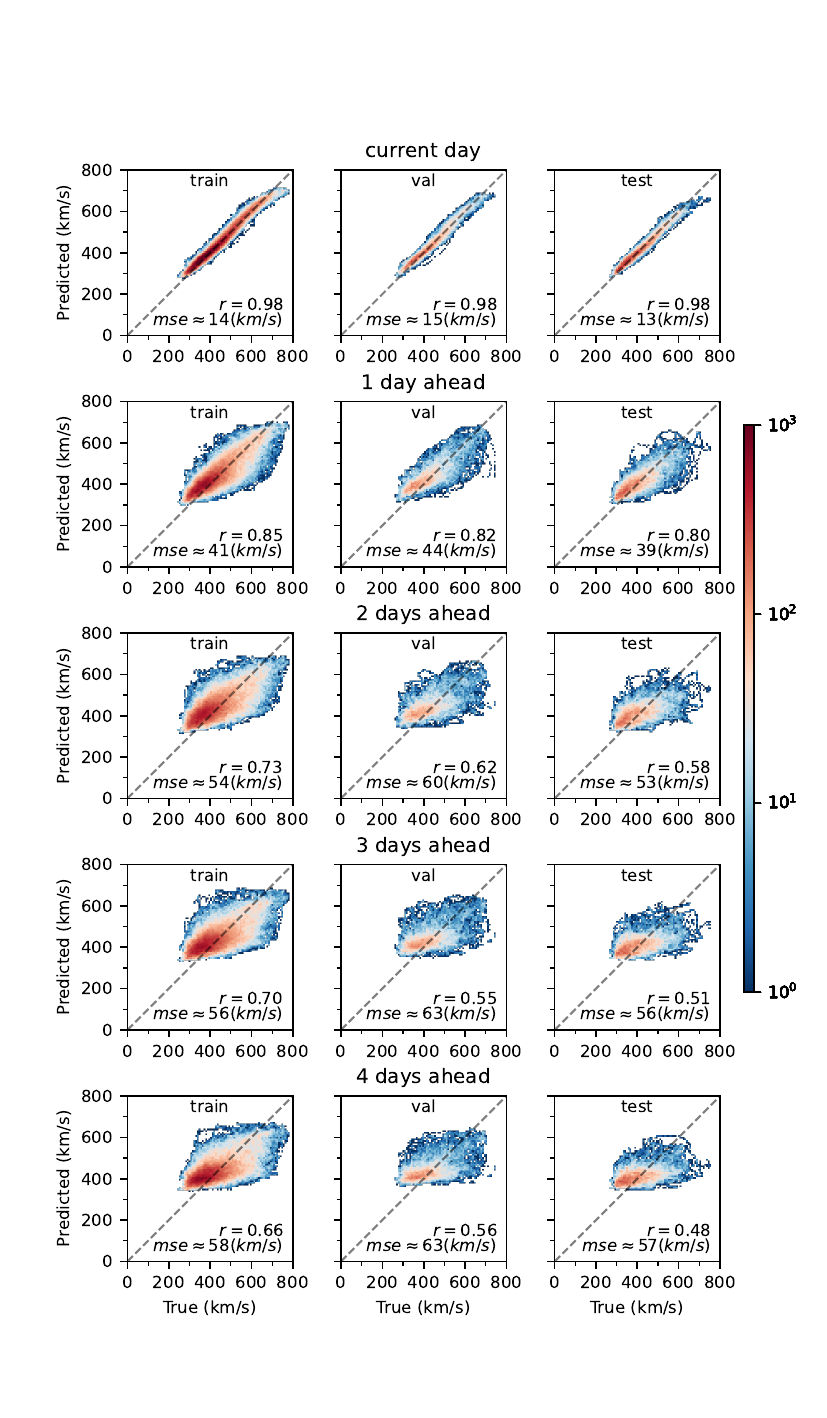}
    \caption{{\bf Heatmaps for true and predicted solar wind speed obtained from the Stage 1 model.} We show heatmaps for the current and future four-day speed predictions. Reconstructions for the current day show an almost-perfect match with the true speed values. Errors for current-day speed reconstructions are concentrated around the extreme true-speed values. For future days, the true and predicted values progressively show further deviations particularly at extreme values $>$ 600 km/s. The shape of the heatmap is also tilted towards the true values, showing that for future two day predictions onward, the higher speed values $>$ 500 km/s are systematically under-predicted.}
    \label{fig:M2Results}
\end{figure*}

\begin{figure*}
    \centering
    \includegraphics[width=0.75\textwidth,trim={1cm 2.1cm 0.2cm 2cm},clip]{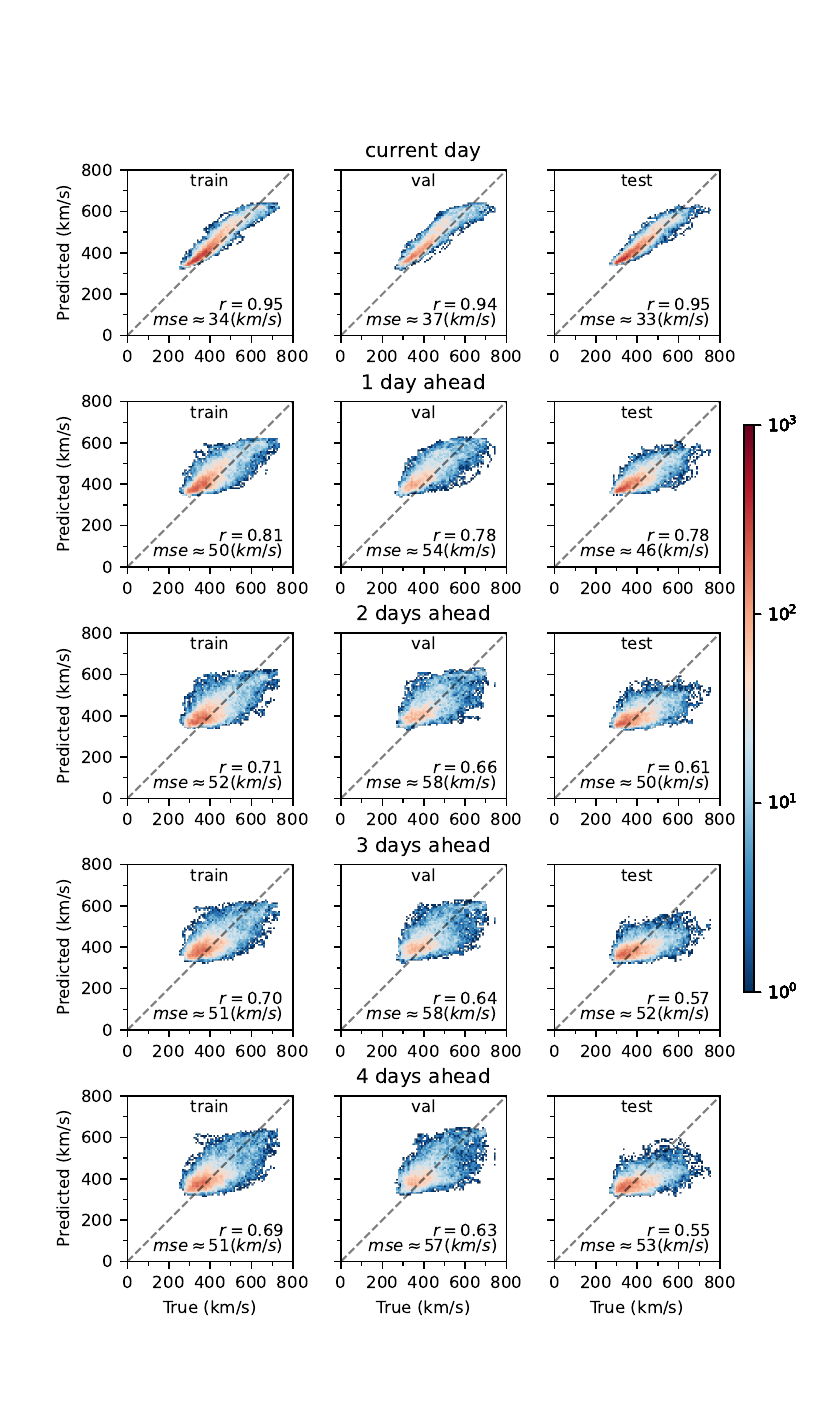}
    \caption{{\bf Heatmaps for the true and predicted solar wind speed obtained from the Stage 2 model.} The heatmaps for the current day reconstruction and future four day speed predictions are shown. The heatmap distributions are qualitatively similar to the Stage 1 model with a quantitative improvement in the reproduction of the higher speed values $>$ 500 km/s.}
    \label{fig:M23Results}
\end{figure*}

\begin{figure*}
    \centering
    \includegraphics[width=0.8\textwidth]{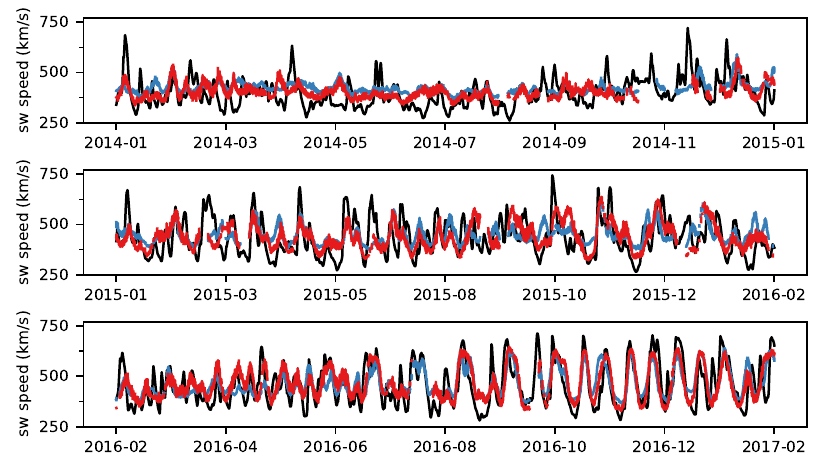}
    \caption{{\bf Comparison of the true and predicted solar wind speeds for the validation data.} The true speeds are denoted in black, while the blue and the red curves respectively denote the predictions from the Stage 1 and Stage 2 model.}
    \label{fig:seriesVal}
\end{figure*}

\begin{figure*}
    \centering
    \includegraphics[width=\textwidth]{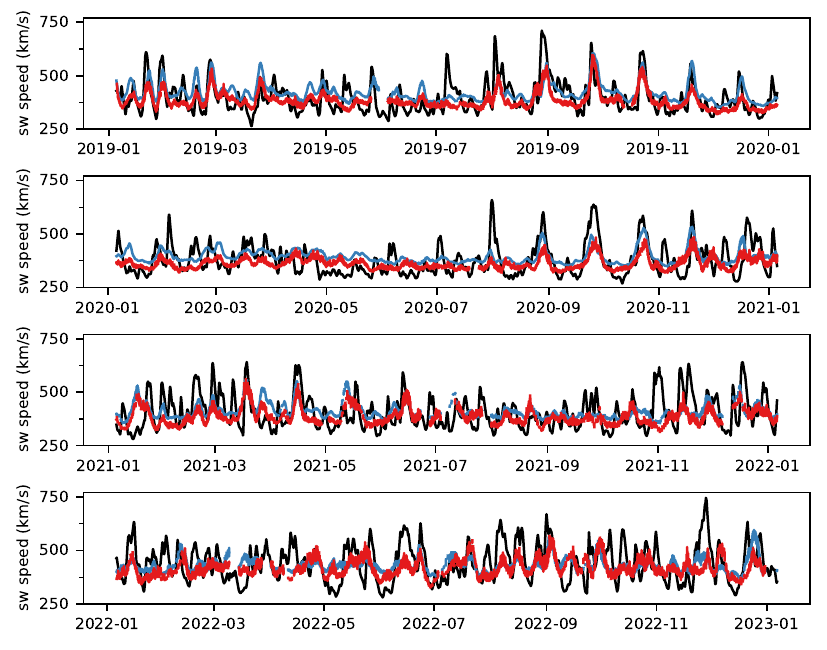}
    \caption{{\bf Comparison of the true and predicted solar wind speeds for the test data.} The true speeds are denoted in black, while the blue and the red curves respectively denote the predictions from the Stage 1 and Stage 2 model. }
    \label{fig:seriesTest}
\end{figure*}
\subsection{Performance comparison}
We compare the r and RMSE values for our model's performance with those of\citet{Upendran2020} and \citet{Brown2022} since these works also use AIA \mathtx{211\AA} images as input. Moreover, \citet{Brown2022} also used the solar wind speed 27 day prior to the fourth day in future as an additional input. However, both of these works used a five-fold cross-validation data split for obtaining the model performance with the validation sets taken from different phases of solar cycle 24. Our work, in contrast, uses a validation set from a period of solar cycle 24 selected such that it includes both the peak and the declining solar activity. The r and RMSE values are expected to be sensitive to the phase of the solar cycle from which the validation data is chosen and therefore a comparison with these past works is only indicative. A more pertinent comparison is obtained by comparing the performance of the model with a one-day and 27-day persistence baseline \citep{Owens2013}. 

A comparison of r and RMSE values from \citet{Upendran2020} and \citet{Brown2022} is presented in Table~\ref{tab:perform2}. It should be noted that \citet{Upendran2020} used four different models to obtain predictions for each future day, while \citet{Brown2022} issue predictions for only the fourth day in the future. The encoder-decoder model in this work yields solar wind speed predictions for all four future days simultaneously (along with the past values). For the fourth day in the future, our model after Stage 2 (that is, after incorporating the AIA images) shows an improved r value of 17\% compared to \citet{Upendran2020} at 0.63. This r value is identical to that of \citet{Brown2022}. The corresponding RMSE value for our model 57.6 km/s is significantly lower than these previous works, by 29\% and 20\% respectively.

Figure~\ref{fig:resultsCorrVar} shows a comparison of the r and RMSE values, for the fourth day of future prediction, with the baseline models of one-day and 27-day persistence, across Solar Cycle 24 and the rising phase of Solar Cycle 25. The periods that form the training, validation, and test sets are marked. In general, the r values for our model's predictions from Stage 1 and Stage 2 are higher than or equal to the 27-day persistence model. The r values for the one-day persistence baseline are occasionally matched by the Stage 2 model. All such instances are during a relatively quiet phase of the cycle. For the validation and test data, the r values from Stage 2 are higher (albeit only marginally at certain phases) than the 27-day persistence baseline, while those from Stage 1 are approximately equal. RMSE values for both the Stage 1 and Stage 2 models are consistently lower than the 27-day baseline model across the training, validation, and test data. As the baseline r and RMSE scores show, the test data in this case are characterized by lower r values and increasing RMSEs compared to the validation data, which also explains the lower r values produced by the Stage 1 and Stage 2 models for the test set. However, the RMSE scores for our models demonstrate a definite improvement over the 27-day persistence baseline, as well as the models of \citep{Upendran2020} and \citep{Brown2022}.

\subsection{Comparison of true and predicted solar wind speed}
The r and RMSE values provide an overall picture, but direct comparison of the true and predicted values gives a clearer understanding of the capabilities and limitations of the model. Heatmaps of the true and predicted values are a straightforward mode of comparing the true and predicted values across all the observed solar-wind speed range. Further, a comparison of the time series of the true and predicted solar wind speeds is useful for understanding the model's performance across the solar cycle and identifying why the model is successful or where it fails. A time series comparison is also useful to understand the scope of the model for operational forecasts. We therefore consider a direct comparison of the true and predicted speeds using heatmaps and time series.

Heatmaps of the true and predicted solar wind speeds obtained from the Stage 1 and Stage 2 models are shown in Figures~\ref{fig:M2Results} and \ref{fig:M23Results}. The heatmaps represent a distribution of the true and predicted values on a scatter plot representing the population in each bin. The points close to the diagonal are most accurately reproduced.  In general, the heatmaps show the highest density of points ($<$ 500 km/s) close to the diagonal, indicating an accurate prediction for most solar wind speed values (and hence relatively higher r values and lower RMSEs). In all heatmaps, including the current day heatmaps, the predictions for the extreme true solar wind speeds $> 600$ km/s are systematically lower, causing a tilt towards the true speed (axis). As forward-looking time increases, the tilt also increases with predictions for true speeds $> 500$ km/s becoming increasingly lower. The width of the distribution about the diagonal also increases, showing a greater spread in the predicted values for a given true value. The Stage 1 and Stage 2 models show qualitatively similar heatmap distributions, although the latter yields improved performance due to a lower deviation and lower spread about the diagonal.

Figures~\ref{fig:seriesVal} and \ref{fig:seriesTest} show the true and predicted values of the validation and test data for the fourth day in future in a time series. The validation data range from the maximum of solar cycle 24 up to a declining phase, and the test data range from the beginning of solar cycle 25 and cover its rising phase. Both the Stage 1 and Stage 2 models predict the speed values accurately during the periods of low solar activity from both cycles. This is expected since these periods are dominated by the CIRs and the consequent long-term correlations in the solar-wind speed variation that are the basis of our encoder-decoder formulation. Note that the extreme speed values as well as the phase are accurately reproduced in both the Stage 1 and Stage 2 predictions during this phase. During the periods of higher solar activity in both datasets, we see that the reproduction of the extreme values is particularly inferior. The extreme solar wind speed values during this phase are primarily due to increased CME rate and hence are not yet captured accurately by the model. However, the Stage 2 model shows a relatively consistent phase match with the true values.

\section{Discussion and Summary \label{sec:discussion}}
We have presented a first encoder-decoder model for forecasting solar wind speed simultaneously for four consecutive days into the future. Our formulation is motivated by the long-lasting CIRs and the resulting 27-day persistence correlation of the solar wind. We used the history of solar wind speed variation from the past solar rotations and the current EUV observations of the Sun in the SDO/AIA channel \mathtx{211\AA}. We used the solar wind speed data from solar cycle 23 and a part of solar cycle 24; and the SDO/AIA data from the same part of solar cycle 24 to train the model. We evaluated the model performance on a validation set from solar cycle 24 consciously chosen to include the period of high and low solar activity. This ensured that the model validation is not biased due to a naturally higher correlation of the 27-day persistence baseline during the solar minimum phase. Our model training is thus also set up for an evaluation of the operational forecasting capabilities, and we demonstrated the model performance on the data from the rising phase of solar cycle 25. Once trained, the model inference is extremely fast, taking $\sim$ 10 ms on a modern desktop computer e.g. a single core of Intel(R) Xeon(R) CPU E5-2620 v4 @ 2.10GHz.

We find that the encoder-decoder model yields a much improved RMSE (by 20\%) for the prediction of the fourth day in the future compared to the state-of-the-art results from \citet{Brown2022}. The Pearson correlation r for the validation data in our case is comparable to \citet{Brown2022}. Our performance generalizes to the test data from solar cycle 25 in terms of RMSE. We also show that our prediction results are consistently superior to the 27-day persistence baseline \citep{Owens2013} across different phases of the solar cycle. The time-series comparison shows that our model reproduces the solar wind speeds accurately during the low solar activity phase dominated by CIRs. The extreme values during the high solar activity phase are systematically underestimated. \citet{Jian2015} benchmarked performance of multiple physics-based semiempirical models used for operational forecasting reporting a correlation of 57\% and $\sim$ 100 km/s for a prediction four days into the future. Compared to these physics-based models \citet{Jian2015} both our r and RMSE values seem to be significantly improved. Note that \citet{Jian2015} evaluated these models on a quiet phase of the solar cycle. As expected, our predictions of the solar wind speed improve as the forward-looking time is reduced and are notably superior to the one-day persistence forecast for the forward-looking time of one day. 

This is a first model featuring an encoder-decoder sequence-to-sequence formulation of the solar wind speed forecasting problem and many improvements are possible. Firstly, the input features considered for the solar wind history are ``hand-picked" and the selection can be improved by surveying other possible important features (e.g. solar wind temperature). In addition, the input features can be further refined after evaluating the feature importance using, e.g., Shapely Values \citep{NIPS2017_7062}. The present multi-modal model exhibits a slight overfitting in training as apparent in Figure~\ref{fig:Losses}. A more efficient fusion of the two modalities, e.g. using early fusion and aligning of the representations \citep{li2021}, may lead to an improved training and therefore improved performance. Also, the model yields the lowest correlation and highest RMSE for solar wind speed predictions during the active phase of the solar cycle. Including other modes of data, such as features from a PFSS magnetic field model as used by \citet{Yang2018,Lin2024} and other channels of SDO/AIA (such as \mathtx{193\AA} considered in the past works) is likely to help improve the model performance by supplementing the information necessary to model the solar wind speed variability and therefore accurately predicting very high speeds $>$ 500 km/sec during the high solar activity phase. Deep learning models are empirical, and hence become more accurate and reliable with a greater amount of data used for training them. However, including different modes of solar observations from different eras poses a challenge of efficiently accounting for the missing-mode observations during the training. Here, this was resolved using a transfer learning approach by incorporating a two-stage training. The problem of missing modes may be handled more efficiently using an elegant formulation designed to model the relevant information of the missing mode of observation from other available modes of the input data. All of these improvements are deferred to follow-up studies. Nevertheless, we hope that the successful encoder-decoder model developed in this work will provide a new direction for deep learning-based reliable operational solar wind forecasting models.

\section*{Acknowledgements}
This work was supported by the New York University Abu Dhabi
(NYUAD) Institute Research Grants G1502 and CG014. This work utilized the High
Performance Computing (HPC) resources of NYUAD ({\it Jubail}). This work also used CPU resources from the Intel(R) vLab.  We acknowledge Nitya Hariharan from Intel Technology India Pvt. Ltd. for suggestions that helped optimize computations on the vLab cluster. This research was also supported in part by a generous donation (from the Murty Trust) aimed at enabling advances in astrophysics through the use of machine learning. Murty Trust, an initiative of the Murty Foundation, is a not-for-profit organization dedicated to the preservation and celebration of culture, science, and knowledge systems born out of India. The Murty Trust is headed by Mrs. Sudha Murty and Mr. Rohan Murty. D.B.D and H.J. acknowledge discussion with Meng Jin from LMSAL/SETI Institute.

\section*{Author Contributions and Data Availability}
D.B.D, S.M.H., and D.D. designed the research and interpreted the results with the help of B.K. D.B.D performed the data analysis with contributions from H.J. and G.S.M. D.B.D. wrote the paper with contributions from S.M.H. and B.K. 

Solar-wind data is publicly available on the OMNIWb website \url{https://omniweb.gsfc.nasa.gov/}. The SDOML/AIA data are publicly available and accessible via \url{https://registry.opendata.aws/sdoml-fdl/}. The sunspot number data are publicly available at SILSO \url{https://www.sidc.be/SILSO/home}. The GOES data are also publicly available and accessible via SunPy \citep{sunpy}. The pre-trained GoogleNet inception model is available via PyTorch.




\end{document}